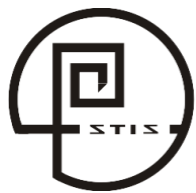

# OSTIS-2015

(Open Semantic Technologies for Intelligent Systems)

# ИССЛЕДОВАНИЕ ЭФФЕКТИВНОСТИ ГЕНЕТИЧЕСКОГО АЛГОРИТМА ДЛЯ ТЕМАТИЧЕСКОГО ДОКУМЕНТАЛЬНОГО ПОИСКА

Иванов В.К., Палюх Б.В.

*Тверской государственный технический университет, г. Тверь, Россия*

**mtivk@tstu.tver.ru**

**pboris@tstu.tver.ru**

В статье приведены результаты экспериментальных исследований эффективности генетического алгоритма, примененного для формирования эффективных поисковых запросов и отбора релевантных результатов при выполнении документального тематического поиска. Исследования проводились с целью сравнительного анализа семантической релевантности и качества ранжирования документов, найденных в Интернет различными способами. Показано, что разработанная технология расширяет возможности семантического поиска и увеличивает число релевантных результатов.

**Ключевые слова:** генетический алгоритм; поисковый запрос; ранжирование; релевантность.

## Введение

Выполнение тематического поиска в документальных хранилищах [Голицына, 2009], является известной процедурой. Однако, как показывает практика, ее эффективное выполнение продолжает оставаться не такой простой задачей, как может показаться (см., например [Попов, 2009]). Одна из основных сложностей – семантически корректная формулировка поисковых запросов, обеспечивающих приемлемые показатели точности поиска. Актуальность разработок новых подходов в этом направлении очевидна, учитывая широкую область применения тематического документального поиска.

В статье представлены результаты экспериментальных исследований эффективности генетического алгоритма, примененного для формирования эффективных поисковых запросов и отбора релевантных результатов поиска. Исследования проводились с целью сравнительного анализа семантической релевантности и качества ранжирования документов, найденных в Интернет различными способами.

## 1. Особенности тематического документального поиска

Тематический документальный поиск ставит целью отбор документов, содержащих координированную информацию (взаимосвязанные факты, их ретроспективу и перспективу) в тематическом сегменте или по заданному объекту. Результатом такого поиска является множество документов, максимально релевантных заданной тематике в целом, а не просто сведения об отдельно взятых событиях, предметах или явлениях. Области применения тематического поиска довольно многообразны: поиск инновационных решений, определение новых направлений бизнеса, сбор информации о клиентах, конкурентные анализ и разведка, обзоры источников научно-технической информации, работа конкурсных комиссий и экспертиза проектов, поиск в патентных исследованиях, подбор учебных материалов.

При выполнении тематического поиска пользователи неизбежно сталкиваются со следующими объективными проблемами:

- Сложность подбора ключевых понятий для формулировки поисковых запросов; невозможность использования в запросах всех доступных критериев отбора документов одновременно.
- Ограничения поисковой системы по составу, структуре и сложности запросов.
- Разрозненность и неоднородность сведений, часто находящихся на стыке смежных областей; наличие альтернатив с сопоставимой релевантностью.
- Отсутствие эффективных систем кластеризации и классификации информации по направлениям тематического поиска.

Решая указанные проблемы, следует правильно интерпретировать результаты поиска, имея в виду совместную оценку релевантности документов, найденных разными запросами, корректность ранжирования поисковой системы, доступность для оценки всех релевантных результатов, наличие эффективных решений в других областях для успешного использования в данной области.

## 2. Реализация генетического алгоритма для фильтрации и ранжирования результатов поиска

В проекте интеллектуальной распределенной системы информационной поддержки инноваций в науке и образовании [Палюх и др., 2013], [Ivanov et al, 2014] предлагается технология генерации поисковых запросов, фильтрации и ранжирования результатов поиска. Эта технология должна быть использована для создания хранилищ инновационных решений образовательного и научного назначения.

Основная идея генерации поисковых запросов – организация с помощью специального генетического алгоритма эволюционного процесса, формирующего в поисковой системе устойчивую и эффективную популяцию запросов для получения высокорелевантных результатов. В ходе этого процесса кодированные запросы, последовательно подвергаются генетическим изменениям и выполняются в поисковой системе. Далее оценивается семантическая релевантность промежуточных результатов поиска, вычисляются значения целевой функции и осуществляется отбор наиболее пригодных запросов.

В [Иванов и др., 2014] представлен прототип программной реализации генетического алгоритма для формирования эффективных поисковых запросов и отбора релевантных результатов при выполнении документального тематического поиска. В частности, описаны основные шаги и параметры алгоритма, компоненты программного обеспечения и результаты предварительных исследований алгоритма. Этот прототип реализован в виде программного продукта Genetic Algorithm Framework (GAF).

В исследовании, описанном в настоящей статье, для проведения экспериментов GAF использовался в следующем составе:

- Пользовательский интерфейс.
- Основная библиотека GAF.
- Модуль морфологического анализа и лемматизации.
- Модуль семантического анализа сходства текстов.
- Модуль поиска для Bing.
- Модуль управления базой данных.
- Модуль управления метаданными

## 3. Подготовка исходных данных

Для участия в исследованиях были приглашены 12 экспертов-исследователей в следующих областях знаний: электронное обучение, базы данных, строительные материалы и изделия, трение и износ в машинах, автоматизация химико-технологических процессов, технологии и средства механизации и технического обслуживания в сельском хозяйстве, медицина, астрофизика.

Каждый эксперт предоставил текстовый материал (автореферат кандидатской или докторской диссертации, монографию, одну или несколько научных статей), наиболее адекватно отражающий ту часть области знаний, в которой он (эксперт) является специалистом.

## 4. Поисковые запросы и параметры их выполнения

На основе материалов, представленных экспертами, были подготовлены запросы двух типов:

- Тип 1. Набор ключевых слов из заглавия предоставленного материала. Длина запроса - от 5 до 11 термов.
- Тип 2. Набор ключевых слов из текста предоставленного материала. Длина запроса - 50 термов. В качестве термов запроса использовались слова материала, имеющие наибольшие веса, вычисленные в соответствие с мерой TF*IDF, но без второго компонента (обратной частоты документа).

Запросы обоих типов формулировались в двух вариантах:

- Вариант 1. Запрос с фиксированными термами (термы в кавычках).
- Вариант 2. Запрос с использованием лемм термов с возможностью словоформ.

Запрос типа 1 выполнялся в поисковой системе Bing со стандартными настройками. Запрос типа 2 выполнялся в среде GAF с модулем поиска для Bing (далее среда GAF/Bing). Идентификаторы запросов, используемые далее в статье, представлены в табл. 1. Отметим, что квалификатор Q (quoted) используется для запросов с вариантом формулировки 1 (термы в кавычках).

Таблица 1 – Идентификаторы типов запросов

| Идентификатор типа запроса | Тип запроса | Вариант формулировки |
|---|---|---|
| Bing.Q | 1 | 1 |
| GAF/Bing.Q | 2 | 1 |
| Bing | 1 | 2 |
| GAF/Bing | 2 | 2 |

Используемые параметры GAF/Bing:

- Количество запросов в генерируемых популяциях $g_2 = 8$.
- Количество ключевых слов в каждом генерируемом запросе $g_3 = 6$.

- Количество результатов поиска, возвращаемых запросом $f_1 = 20$, либо популяцией запросов $f_2 = 20$, либо суммарно всеми популяциями $f_3 = 20$.
- Коэффициент для учета расположения документов на одном сервере $f_4 = 0{,}75$.
- Весовые коэффициенты для аргументов $p$, $r$ и $s$ соответственно при расчете ранга результата поиска $f_5 = 0{,}33$, $f_6 = 0{,}33$, $f_7 = 0{,}34$ (о функции пригодности и ее аргументах см. далее раздел "Сравнительный анализ качества ранжирования").
- Вероятность мутации запроса $m_1 = 1$.
- Число проходов алгоритма (или число генерируемых популяций) $e_1 = 10$.

## 5. Обработка результатов экспертами

Результаты выполнения всех запросов по каждой тематике (списки адресов документов в Интернет, упорядоченные по неизвестному экспертам алгоритму) передавались экспертам для оценки их релевантности.

Под релевантностью предлагалось понимать соответствие содержания материала теме исследований, проведенных или проводимых экспертом. Должно было оцениваться соотношение объёма информации, полезной для разработки темы, к общему объёму информации в материале. Для оценки релевантности документа предлагалось использовать следующую шкалу:

- 0 – материал не содержит информации, полезной для разработки темы.
- 1 – материал содержит незначительное количество информации, полезной для разработки темы.
- 2 – материал содержит достаточное количество информации, полезной для разработки темы.
- 3 – материал содержит значительное количество информации, полезной для разработки темы.

Оценку каждого найденного документа следовало выполнить:

- С позиций высококвалифицированного специалиста по теме исследований эксперта (релевантность специалиста).
- С позиций специалиста, только *начинающего* научную работу по теме исследований эксперта (релевантность новичка).

Экспертам рекомендовалось учитывать, что полезность найденных документов может определяться похожими исследованиями, справочными данными, ссылками на другие материалы, описаниями аналогичных применяемых моделей, методов и технологий. При сомнениях в выборе одного из двух значений баллов шкалы предлагалось использовать меньшее значение. Также, релевантность информационных материалов из справочников, словарей, энциклопедий рекомендовалось, как правило, оценивать баллами 0 или 1 (исключения – уникальная справочная информация).

Не рекомендовалось рассматривать как полезные рекламные объявления, коммерческие предложения, "визитные карточки" предприятий и т.п.

Для сохранения всех исходных данных и результатов исследований была разработана специализированная база данных.

## 6. Сравнительный анализ релевантности результатов поиска

На рис. 1 и 2 показана средняя релевантность документов – результатов поиска, определенная экспертами для специалистов и новичков соответственно.

В случае использования в запросах лемм термов с возможностью словоформ средняя релевантность документов, найденных Bing, на 3…9% выше, чем у документов, отобранных GAF/Bing. Причем это касается релевантности, определенной как для специалистов, так и для новичков. Для запросов с фиксированными термами Bing дает еще лучшую (на 11…13%) релевантность обоих типов для отобранных документов.

В целом, соотношения средней релевантности документов для специалистов и новичков для различных типов запросов сохраняются.

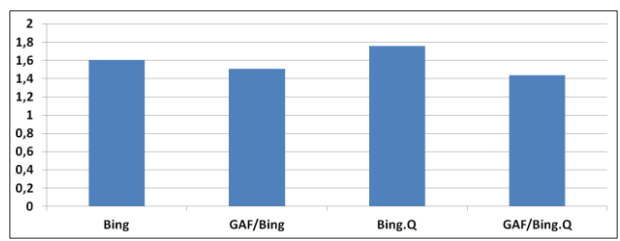

Рисунок 1 – Средняя релевантность результатов поиска (для специалистов)

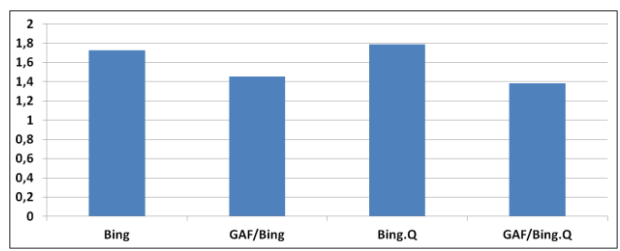

Рисунок 2 – Средняя релевантность результатов поиска (для новичков)

На рис. 3 и 4 показана точность поиска - доля релевантных документов в результатах поиска, определенная экспертами для специалистов и новичков соответственно.

В случае использования в запросах лемм термов доля релевантных результатов для специалистов в результатах поиска выше для Bing на 5%. Для новичков доля релевантных результатов выше для

GAF/Bing на 7%. Для запросов с фиксированными термами доля релевантных результатов для специалистов примерно одинакова для Bing и для GAF/Bing, но для новичков доля релевантных документов, отобранных GAF/Bing выше на 4%.

В целом можно отметить более высокую долю релевантных документов, найденных с применением GAF/Bing.

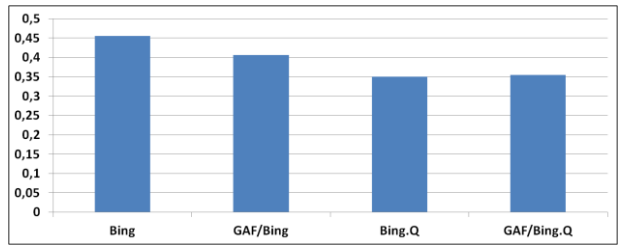

Рисунок 3 – Точность поиска - доля релевантных результатов в результатах поиска (для специалистов)

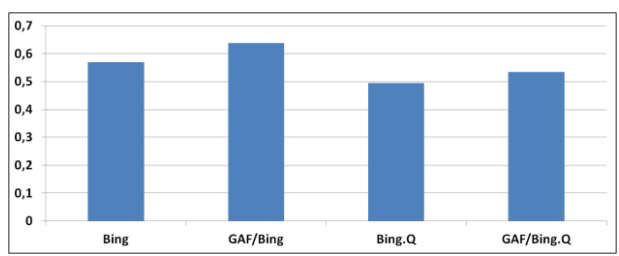

Рисунок 4 – Точность поиска - доля релевантных результатов в результатах поиска (для новичков)

## 7. Сравнительный анализ качества ранжирования

Ранжирование результатов поиска в GAF осуществляется в ходе вычисления значений функции пригодности $W$ при выполнении генетического алгоритма. Значение $W$ определяет качество запросов (пригодность особей популяции); генетический алгоритм ищет максимум $W$:

$$\overline{W} = \frac{1}{N}\sum_{j=1}^{N}\overline{w}_j \to \max \qquad (1)$$

где
$W$ – функция пригодности популяции;
$N$ – количество запросов в популяции;
$w_j$ – функция пригодности $j$-го запроса популяции.

$$\overline{w} = \frac{1}{P}\sum_{i=1}^{P}w_i(f,p,s,a) \qquad (2)$$

где
$P$ – количество результатов поиска в запросе;
$w_i$ – функция пригодности $i$-го результата запроса.

Для ранжирования результатов поиска используется значение $w_i(f, p, s, a)$. Для каждого $i$-го результата запроса значение $w_i$ зависит от следующих аргументов: $f$ – определяется позицией результата в ранжированном списке результатов, построенном используемой поисковой системой;

$p$ – определяется вхождением данного результата в списки результатов большинства запросов; $s$ – определяется семантической близостью к эталонным текстам, адаптивно формируемым в ходе выполнения алгоритма; $a$ – определяется параметрами пользователя и является фактором среды (значения $f, p, s, a$ нормированы на диапазон от 0 до 1). Для ранжирования использовалась средневзвешенная сумма $f, p, s,$ и $a$. Отметим, что предлагаемый способ ранжирования уточняет подходы, принятые в поисковых системах, которые, в свою очередь, основаны на сочетании в разной степени семантической релевантности и авторитетности документов, а также на поведенческих особенностях пользователей.

Для оценки качества ранжирования результатов поиска использовались метрики DCG@n и NDCG@n [Järvelin et al., 2002], [Агеев и др., 2010]. Эти метрики оценивают качество первых $n$ документов в результатах поиска. Для документов, упорядоченных по значению функции пригодности $w_i$, вычислялись значения

$$DCG = \frac{1}{N}\sum_{p=1}^{P}\frac{2^{gr(p)}-1}{\log_2(2+p)} \qquad (3)$$

где
$gr(p)$ – средняя экспертная оценка релевантности, выставленная документу, расположенному на позиции $p$ в списке результатов;
$gr \in [0,3]$, причем 3 означает "релевантный", 0 – "нерелевантный", 1 и 2 – "частично релевантный": "релевантный(+)" или "релевантный(–)";
$1/\log_2(2+p)$ – дисконт за позицию документа (документы в начале списка имеют больший вес).

На рис. 5 показано соотношение значений метрик $DCG$ для идеального ранжирования, то есть ранжирования, выполненного экспертами (DCG*Expert), ранжирования, выполненного поисковой системой (DCG*Bing), и ранжирования, выполненного GAF (DCG*GAF/Bing).

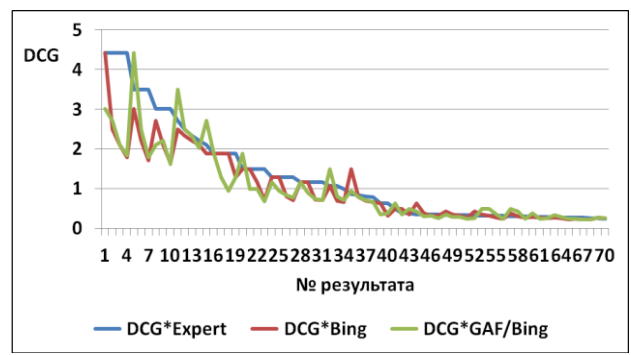

Рисунок 5 – Значения метрики DCG релевантных результатов поиска (для специалистов)

В экспериментах использовались запросы Bing (см. табл. 1) со стандартными параметрами поисковой системы Bing и запросы GAF/Bing (см. табл. 1) с параметрами: $g_2 = 1$, $g_3$ – соответствует запросу Bing, $f_1 = f_2 = f_3 = 20$, $f_4 = 0{,}75$, $f_5 = 0{,}33$, $f_6 = 0{,}33$, $f_7 = 0{,}34$, $m_1 = 1$, $e_1 = 10$. Описание параметров приведено выше в разделе "Поисковые запросы и параметры их выполнения".

В целом графики на рис. 5 показывают хорошее визуальное совпадение значений *DCG* для рассматриваемых способов ранжирования.

Для точной количественной оценки сходства значений из пар последовательностей ρExBing = (DCG*Expert, DCG*Bing) и ρExGAF = (DCG*Expert, DCG*GAF/Bing) были рассчитаны значения коэффициента взаимной корреляции $\rho_{12}$:

$$\rho_{12}(j) = \frac{r_{12}(j)}{\frac{1}{N}\left[\sum_{n=0}^{N-1} x_1^2(n) \sum_{n=0}^{N-1} x_2^2(n)\right]^{1/2}} \quad (4)$$

где

$\rho_{12}$ – коэффициент взаимной корреляции значений из последовательностей 1 и 2;

$j$ – сдвиг значения, $j=0$ в рассматриваемом случае;

$n$ – номер значения в последовательностях;

$N$ – число значений в каждой последовательности;

$x_1$, $x_2$ – значения из последовательностей 1 и 2, в рассматриваемом случае это значения из DCG*Expert и значения из DCG*Bing либо из DCG*GAF/Bing;

$r_{12}$ – взаимная корреляция значений из последовательностей 1 и 2:

$$r_{12}(j) = \frac{1}{N}\sum_{n=0}^{N-1} x_1(n) x_2(n+j) \quad (5)$$

При проведении расчетов использовалась методика из [Айфичер, 2004].

На рис. 6 представлена сравнительная оценка значений $\rho_{12}$, рассчитанных для значений из пар ρExBing и ρExGAF. Принимались во внимание оценки релевантности для специалистов и новичков. Диаграммы показывают, что качество ранжирования Bing незначительно выше, чем дает GAF/ Bing - значение $\rho_{12}$ для пары ρExBing выше на 1,3...1,4%, чем значение $\rho_{12}$ для пары ρExGAF. При оценке релевантности результатов для новичков значения $\rho_{12}$ выше и для ρExBing и для ρExGAF.

Далее для результатов поиска Bing и GAF/Bing вычислялись нормализованные значения $NDCG = \frac{DCG}{Z}$, где $Z$ –максимально возможное значение *DCG* для случая идеального ранжирования в соответствие с оценками эксперта. Показатель *NDCG* принимает значения от 0 до 1.

Соотношение значений *NDCG* для результатов поиска Bing (NDCG*Bing) и GAF/Bing (NDCG*GAF) с детализацией по релевантности для специалистов и новичков представлено на рис. 7.

Видно, что нормализованная метрика *DCG* в целом дает оценку качества ранжирования GAF/Bing выше (0,8%), чем ранжирования Bing. Особенно это проявляется при оценке релевантности документов с позиций новичков.

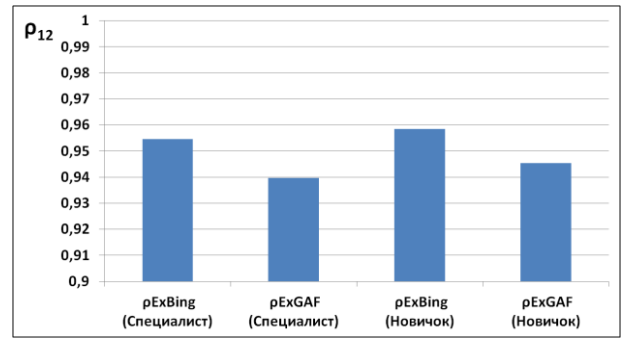

Рисунок 6 –Коэффициенты взаимной корреляции значений метрик *DCG* релевантных результатов поиска (для специалистов и новичков)

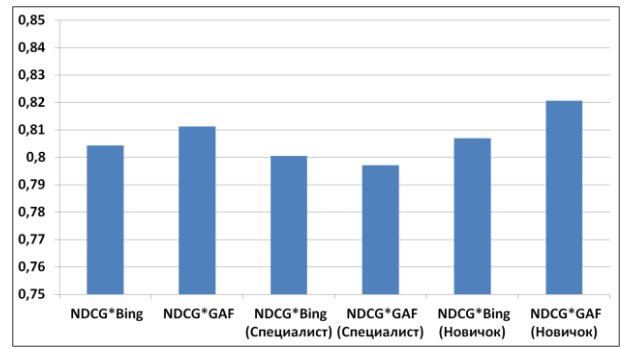

Рисунок 7 – Нормализованные значения метрики *DCG*

## 8. Выводы по результатам экспериментов

1. Точность поиска или доля релевантных результатов, найденных GAF/Bing, выше, чем у Bing в случае определения релевантности для новичков. Отметим, что точность поиска является очевидным наиболее значимым критерием оценки результатов поиска.

2. Процент совпадений результатов (адресов найденных документов) при поиске через GAF/Bing и Bing – 2%. Очевидно, что технология GAF/Bing позволяет находить новые оригинальные документы.

3. Средняя релевантность документов, найденных с использованием предлагаемой технологии GAF/Bing, практически не уступает средней релевантности документов, найденных Bing. Причем разрыв уменьшается для запросов, состоящих из лемм с возможностью словоформ.

4. Качество ранжирования результатов поиска GAF/Bing соответствует качеству ранжирования Bing. Значения коэффициента взаимной корреляции *DCG* для ранжирования GAF/Bing и экспертного ранжирования сопоставимы. То же можно сказать о значениях нормализованной метрики *NDCG*.

## Заключение

Анализ результатов исследования

эффективности генетического алгоритма для тематического документального поиска показывает, что наибольший эффект от применения разработанной технологии достигается при поиске источников информации по новой для специалистов теме на начальных этапах ее изучения и освоения. При этом существенно увеличивается число уникальных и достаточно релевантных результатов поиска.

В дальнейших исследованиях авторы предполагают провести тестирование разработанной технологии на дорожках TREC, таких как поиск в Web [Collins-Thompson et al, 2014] и KBA [Frank et al, 2014].

## Библиографический список

# STUDY THE EFFECTIVENESS OF GENETIC ALGORITHM FOR DOCUMENTARY SUBJECT SEARCH

Ivanov V.K., Palyukh B.V.

*Tver State Technical University, Tver, Russia*

mtivk@tstu.tver.ru

pboris@tstu.tver.ru

This article presents results of experimental studies the effectiveness of the genetic algorithm that was applied to effective queries creation and relevant document selection. Studies were carried out to the comparative analysis of the semantic relevance and quality ranking of the documents found on the Internet in various ways. Analysis of the results shows that the greatest effect of presented technology is achieved by finding new documents for skilled users in the initial stages of the study of the topic. Additionally, the number of unique and relevant results is significantly increased.

**Introduction**

One of the main difficulties during the subject search in documentary data warehouses is correct search queries formulating that should provide acceptable accuracies level. This article presents results of experimental studies the effectiveness of the genetic algorithm that was applied to effective queries creation and relevant document selection.

**Main Part**

The genetic algorithm for filtering and ranking the search results has been implemented previously. Source data for the experiments were prepared with twelve experts together.

Next, the initial search queries were formulated, algorithm parameters were assigned and large series of query generations and search operations were performed. The results were processed by involving experts.

In conclusion the comparative analysis the relevance of search results and ranking quality were performed.

**Conclusion**

Analysis of the results shows that the greatest effect of presented technology is achieved by finding new documents for skilled users in the initial stages of the study of the topic. Additionally, the number of unique and relevant results is significantly increased.